\documentstyle[aps,prb,epsf]{revtex}

\def\Journal#1#2#3#4{{#1} {\bf #2}, #3 (#4)}

\def\be{\begin{equation}}
\def\ee{\end{equation}}
\def\bea{\begin{eqnarray}}
\def\eea{\end{eqnarray}}
\def\ve{\varepsilon}

\def\g{\gamma}

\def\Tr{\mbox{Tr}}

\def\nne{\nonumber\\ }
\def\nn{\nonumber}

\def\qab{q^{\alpha\beta}}

\def\qgd{q^{\gamma\delta}}
\def\sia{S^{\alpha}_i}
\def\sib{S^{\beta}_i}
\def\sja{S^{\alpha}_j}

\def\sa{S^{\alpha}}
\def\sb{S^{\beta}}
\def\a{\alpha}
\def\b{\beta}
\def\g{\gamma}
\def\d{\delta}

\def\sumij{\sum_{\langle ij \rangle}}

\def\sumab{\sum_{\alpha < \beta}}
\def\suma{\sum_{\alpha=1}^n}
\def\bc{\begin{center}}
\def\ec{\end{center}}
\def\la{\langle}
\def\ra{\rangle}

\def\a{\alpha}
\def\b{\beta}
\def\g{\gamma}
\def\d{\delta}

\def\ve{\varepsilon}
\def\Tr{\mbox{Tr}}

\def\sa{S^{\alpha}}
\def\sb{S^{\beta}}
\def\sia{S_i^{\a}}
\def\sib{S_i^{\b}}

\def\sja{S_j^{\a}}

\def\sumij{\sum_{\la\!ij\!\ra}}

\def\suma{\sum_{\a}}

\def\sumalb{\sum_{\alpha<\beta}}

\def\xiab{\xi^{\a\b}}
\def\qab{q^{\a\b}}
\def\qbg{q^{\b\g}}
\def\qda{q^{\d\a}}
\def\qga{q^{\g\a}}

\def\qgd{q^{\g\d}}

\def\be{\begin{equation}}
\def\ee{\end{equation}}
\def\bi{\begin{itemize}}
\def\ei{\end{itemize}}
\def\bea{\begin{eqnarray}}
\def\eea{\end{eqnarray}}
\def\nn{\nonumber}
\def\nne{\nonumber\\}
\def\le0{\Biggr|_{\ve=0}}
\def\1n{\frac{1}{N}}
\def\dab{\partial_{\a\b}}
\def\tdab{\tilde{\partial}_{\a\b}}

\begin{document}
\draft
%
\begin{title}
{Mean-field glassy phase of the random field Ising model}
\end{title}
\author{
 A. A. Pastor and V. Dobrosavljevi\'{c}}
\address{Department of Physics and National High Magnetic Field Laboratory\\ Florida State University, Tallahassee, Florida 32306}

\author{ M. L. Horbach$^{\dagger}$}
\address{Serin Physics Laboratory, Rutgers University\\
Piscataway,  NJ 08854-8019}

\maketitle
\begin{abstract}
The emergence of glassy behavior of the random field Ising model
(RFIM) is investigated using an extended mean-field theory
approach. Using this formulation, systematic corrections to the
standard Bragg-Williams theory can be incorporated, leading to the
appearance of a glassy phase, in agreement with the results of the
self-consistent screening theory of Mezard and Young.  Our
approach makes it also possible to obtain information about the
low temperature behavior of this glassy phase. We present results
showing that within our mean-field framework, the hysteresis and
avalanche behavior of the RFIM is characterized by power-law
distributions, in close analogy with recent results obtained for
mean-field spin glass models.
\end{abstract}

%
\newpage

\section{Introduction}

The random-field Ising model (RFIM)\cite{rfim} is one of the simplest
models used to describe the frustration introduced by disorder in
interacting many-body systems. Despite its simplicity, its behavior
proved to be the source of much controversy, primarily due to the lack
of reliable theoretical methods that one can use for such
systems. Still, efforts to elucidate the basic properties of the RFIM
continue to attract considerable attention, primarily because of its
direct relevance to a number of important physical problems. These
include not only the behavior of diluted magnets in external magnetic
fields,\cite{monten} but also several aspects of electronic transport
in disordered insulators\cite{efros} and systems near the
metal-insulator transition.\cite{tedrf,pastor} In addition, the
non-equilibrium behavior of the RFIM has been used to model the
physics of hysteresis and avalanche behavior and the origin of
self-organized criticality\cite{sethna}. Finally,
non-disordered models with infinitesimal random fields have been
studied\cite{ioffe} in order to investigate self-generated glassy
behavior\cite{self-glass} observed in systems such as supercooled
liquids\cite{teddave}, or even under-doped cuprates.\cite{schmalian}

The simplest effect of turning on a weak random field is the
resulting depression of the critical temperature for uniform
ordering, while for sufficiently strong randomness the ordered
phase completely disappears. This behavior is apparent even in the
simplest Bragg-Williams (BW) mean-field theory (MFT), but
understanding the disorder-induced modification of the relevant
critical behavior proved much more difficult. Very early on,
perturbative renormalization group (RG) results of Parisi and
Sourlas\cite{parsur} (PS) suggested the existence of a
``dimensional reduction'' by which the random problem belongs to
the same universality class as a clean one in two dimensions less.

Unfortunately, the beautiful result of PS was found to be in conflict
not only with the heuristic argument of Imry and Ma,\cite{imryma} but
also with the exact results of Imbrie,\cite{imbrie} and thus deemed
incorrect. More recently, the origin of these discrepancies was
traced\cite{mezyoung} to the implicit assumption of PS that outside
the ordered phase, a single thermodynamically stable state exist. This
assumption may not be warranted if another phase transition,
presumably of glassy character, would precede any uniform ordering,
which would lead to the breakdown of dimensional reduction. Of course,
such a glassy phase is not found in the naive BWMFT, so that more
sophisticated theoretical schemes have to be used in order to identify
the corresponding instability of the high temperature paramagnetic
phase. Such a theory was first formulated by Mezard and
Young,\cite{mezyoung} who utilized the self-consistent screening (SCS)
approach of Bray,\cite{bray} and identified the glassy phase by
carrying out a replica-symmetry breaking stability analysis. Similar
results were obtained by numerically solving the mean-field equations
for a fixed realization of disorder by Lancaster et
al.,\cite{lancaster} confirming the existence of the glassy
phase. Finally, Mezard and Monasson,\cite{mezmon} and De Dominicis et
al.\cite{dominicis} presented arguments that the glass phase should
persist even at weak disorder, and everywhere precede the uniform
ordering, in agreement with the breakdown of the dimensional
reduction.

While these approaches provided important new information on the RFIM,
several aspects remained unsatisfactory. The SCS scheme, while being
able to identify the glass phase, proved of considerable complexity,
making it difficult to obtain simple analytical results. In addition,
the physical content of this formulation does not appear very
transparent, making it difficult to establish what crucial ingredients
are necessary for a theory in order to be able to identify and
describe such a glass phase.  Finally, the low temperature properties
of the glass phase also seem very difficult to establish using this
approach, which presents a severe limitation in applying the RFIM to
the problem of the ``Coulomb glass''\cite{pastor} and the related
theory of Efros and Shklovskii.\cite{efros}

The main goal of the present paper is to identify the simplest
possible approach that is capable of providing a description of the
glassy phase.  We will show that to do this, two requirements have to
be satisfied: (1) the theory has to identify the correct order
parameters, and (2) it has to incorporate spatial fluctuations beyond
the naive BW theory.  Both of these are automatically satisfied by the
SCS theory, but we will show how the same goals can be achieved in a
simpler and more transparent fashion, by examining systematic
corrections to the BW theory. To do this, we follow an approach used
by Plefka\cite{plefka} to derive the Thouless-Anderson-Palmer (TAP)
equations for spin glasses. \cite{spinglass} A similar formulation was
subsequently used by Georges et al.,\cite{georges} to obtain
systematic corrections to the MFT of the Sherrington-Kirkpatrick (SK)
model. This approach fixes the desired order parameters by introducing
appropriate source fields, and then evaluates the corresponding Gibbs
free energy by an expansion in powers of the interaction $J$.  A brief
description of some of our results in the electronic context has
already been presented in Ref. [5], but the present paper provides
many more details and a number of new results.

The rest of the paper is organized as follows. In section II, we
begin our discussion by examining the RFIM on a Bethe lattice,
where a particularly simple derivation of the BW theory and its
leading corrections can be obtained, resulting in the emergence of
the glass phase. A more general strategy based on the Plefka
approach is presented in section III, showing that these leading
corrections take the same form on arbitrary lattices. The low
temperature structure of the glass phase is discussed in section
IV, where the hysteresis and avalanche behavior in our extended
MFT is examined, showing the emergence of self-organized
criticality similar to that recently discovered by Pazmandi et
al.\cite{zimanyi} for the SK model. The role of higher order
corrections to our theory and its relation to the $1/z$ expansion
approach is discussed in section V, where we show that the leading
nontrivial corrections to the BWMFT represent dominant
contributions in the joint limit of large coordination {\em and}
large random fields. In this section, we also comment on the
relevance of our approach to controversial question of
self-generated glassy behavior in systems without disorder, which
we discuss by examining the limit of weak random fields.  Our
conclusions are summarized in section VI, where we also outline
some interesting directions for future work.

\section{The Bethe Lattice}

The simplest theoretical formulation of the random field Ising
(RFIM) is obtained by considering the large coordination limit,
where the standard Bragg-Williams mean-field theory (BWMFT)
becomes exact. However, one has to go beyond this limit in order
to obtain nontrivial results, and we address this question by
examining systematic corrections to the BWMFT. The large
coordination limit and the leading corrections are particularly
easily formulated in the special case of the Bethe lattice, where
a simple recursive procedure can be used.  This approach also
gives some insight in the mechanism for the emergence of the glass
phase, so we begin our discussion by concentrating on the Bethe
lattice in the limit of large coordination.

The Hamiltonian of the random-field
Ising model is given by

\bea H_{int}=-\sumij
J_{ij}S_iS_j-\sum_ih_iS_i\label{eq:hamiltonian}.
\eea
Here, $S_i =\pm 1$, and $J_{ij}={J\over z}$ are uniform ferromagnetic
interactions between nearest neighbor sites, rescaled with the
coordination number $z$ in order to obtain a finite result in the
$z\rightarrow \infty$ limit.  The random magnetic fields $h_i$ are
assumed to be Gaussian distributed, with zero mean and a variance
$\langle h_i^2\rangle = H_{RF}^2$.

Using standard replica methods \cite{spinglass}, we can
formally average over disorder, and the resulting
partition function takes the form ($\alpha =1,...,n;\;\; n\rightarrow 0$)
\be Z^n  = \Tr\exp [ {\b J\over z}\suma
\sumij \sa_i\sa_j + \frac{1}{2}(\beta H_{RF})^2 \sum_i(\suma \sa_i)^2
].
\ee
We proceed by taking advantage of the tree-like structure of the Bethe
lattice, by formally summing over all the degrees of freedom in one
branch. The resulting functional $\Phi(\sa_0)$ is a function only of
the variable $\sa_0$ at the branch origin, and can be easily seen to
obey the following self-consistent equation
\be
\Phi(\sa_0)=\Tr_{\sa_1}( \exp [ {\b J\over z} \suma \sa_0 \sa_1
+
\frac{1}{2}(\b
H_{RF})^2(\suma \sa_1 )^2] \Phi^{z-1} (\sa_1)).
\ee

\subsection{Bragg-Williams Theory}

In order to examine the large $z$ limit, it is convenient to define a
single-site effective action by
\be
L[\sa]\equiv -\frac{1}{2}(\b H_{RF})^2(\suma \sa_1 )^2
-\ln(\Phi^{z-1}(\sa)).\ee
In the $z\rightarrow\infty$ limit, this expression
simplifies, since the interaction has been scaled by $1/z$, and
the functional $\Phi(\sa_0)$ can be obtained by expanding
the self-consistency condition, Eq. (3) in powers of the
interaction $J$. To leading order, only the terms {\em linear}
in $J$ survive, and we find
\be
L^{(1)}[\sa]= -\b J\suma\sa m^{\alpha}
-(\b H_{RF})^2\sumab \sa\sb .
\ee
Here, the index $(1)$ indicates that only terms linear in $J$ are retained.
As we expect for a Bragg-Williams theory, this local effective action
corresponds to a single spin which, in addition to the local random
field, also experiences the presence of a ``molecular'' field
$Jm^{\alpha}$ due to interaction with the neighbors.

We emphasize that this procedure {\em automatically} defines the order
parameters entering the local effective action. In the
$z\rightarrow\infty$ limit only the magnetization $m^{\a}$ appears,
which from Eq. (3) satisfies the following self-consistency condition
\bea
m^{\a}= \langle\sa\rangle_{L[S]}.
\eea
Since the interaction term does not mix the replicas, they
trivially decouple, and the self-consistency condition for
the magnetization $m^{\a}=m$ takes the form
\be
m=\int{Dx\tanh((\b H_{RF} x+\b Jm)},
\ee
where $Dx=dx\exp(-x^2/2)/\sqrt{2\pi}$. This equation is a
straightforward generalization of the well-known Bragg-Williams
condition to include the effect of random fields. The critical
temperature where the magnetization vanishes is easily computed, and
is found to vanish at a critical strength $H_{RF}^c /J = \sqrt{2/\pi}$ of
the random fields, as shown in Figure 1(a). Outside this ferromagnetic
phase the local effective action of Eq. (5) reduces to that of
noninteracting spins in the presence random fields, and no further
phase transition can be found.

\begin{figure}
\bc \epsfxsize=4in \epsfbox{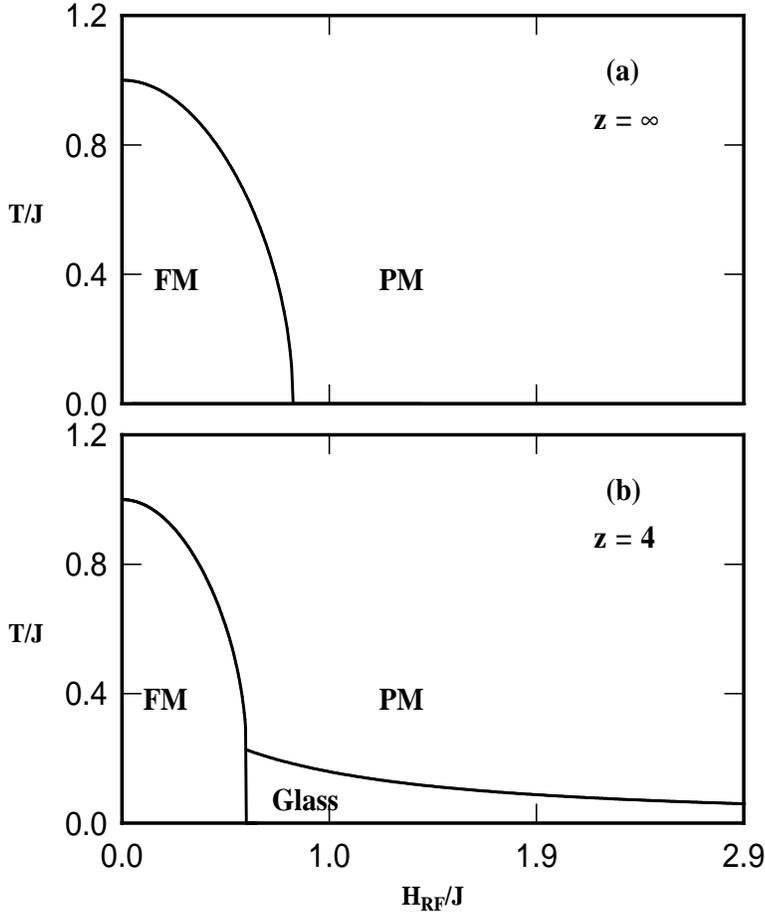}\ec \caption{Phase diagram
for coordination numbers $z=\infty$ (a) and $z=4$ (b).
Ferromagnetic (FM), paramagnetic (PM) and glass phases are found.
Note that the glass phase does not exist for $z=\infty$. }
\end{figure}

The reason for this limitation of the BWMFT is very simple, and can be
appreciated by considering of the (thermally averaged) Weiss field
$h_i^W = <\sum_j J_{ij} S_j >_T$ produced by the neighboring spins on a
given site. In a uniform system, $h_i^W$ is the same on every site, but
in presence of randomness, it may display appreciable spatial
fluctuations. This is especially important outside the any uniformly
ordered phase, where the spatial average $\overline{h} = <h_i^W>_{site}$
vanishes, but the {\em} local value $h_i^W$ of the Weiss field may
remain finite, reflecting the {\em local} breaking of the up-down
symmetry. This behavior is encountered in spin glasses, where
$\overline{h_i^W} =0$, but $\overline{(h_i^W )^2}$ becomes finite
below a temperature corresponding to the glassy freezing.
In the simple BWMFT, only the first moment $\overline{h_i^W} \sim m$
is retained, thus its inability to describe any glassy ordering.

\subsection{Leading corrections and glassy freezing}

In order to search for the existence of nontrivial behavior outside
the uniformly ordered phase, we have to go beyond the BWMFT, i. e. the
$z=\infty$ limit. The leading correction is obtained by iterating the
self-consistency condition Eq. (3) to second order in the interaction
$J$, and in the expression for the local effective action a new term,
quadratic in $J$ appears
\bea
L &=& L^{(1)} + L^{(2)},\nn\\
L^{(2)}&=& -\frac{1}{z}(\b J)^2\sumab\sa\sb\left[\qab-(m^{\alpha})^2\right].
\eea
The order parameter $m^{\alpha}$ is still given by Eq. (6) (although
the average is now computed with respect to the extended action of
Eq. (8)), but the new order parameter $\qab$ appears, which is
self-consistently determined by
\be \qab=\langle\sa\sb\rangle_{L[S]}.\ee
This quantity is nothing but the familiar Edwards-Anderson order
parameter\cite{spinglass}. Note that in our Bethe lattice procedure,
it appears automatically, as a result of an expansion in powers of the
interaction $J$ to the lowest nontrivial order beyond the BWMFT.  Its
presence reflects the fact that in finite-coordinated lattices, local
Weiss fields $h_i$ are random numbers with a finite dispersion. More
precisely, they are described by the distribution of local fields
$P(h_i)$, the explicit form of which is determined by the replica matrix
$\qab$.

As in standard spin-glass theory\cite{spinglass}, the solution of
these equations assumes the simplest form in the high temperature
phase, where replica symmetry is valid, such that $m^{\alpha}=m$ and
$\qab = q$.  In this case, our self-consistency conditions take the
form
\bea
m&=&\int{Dx\tanh\left[ \left((\b H_{RF})^2 +{(\b J)^2\over
z}(q-m^2)\right)^{1/2} x+\b Jm\right] }\nne
q&=&\int{Dx\tanh^2\left[\left((\b H_{RF})^2+{(\b J)^2\over
z}(q-m^2)\right)^{1/2} x+\b Jm\right]}.
\label{eq:qbethe}
\eea
This is sufficient to determine the ferromagnetic (FM) phase boundary
which, as before, is determined by setting $m=0$. By numerically
solving these equations, we find that the FM phase is only slightly
reduced due to the fluctuation corrections, as shown in Figure 1(b).

Identifying the glassy freezing is more difficult. For standard
spin-glass models, the glassy freezing coincides with the breaking of
the up-down symmetry, so that the glass transition temperature can be
identified even within replica symmetric theory, as the point where
the Edwards-Anderson order parameter $q=\overline{<S_i >^2}$ assumes a
nonzero value. In our case, the random magnetic field plays a role of
a source conjugate to the order parameter, locally breaking the
up-down symmetry. The situation is similar as in spin glass models in
a uniform external field \cite{spinglass}, where the replica symmetric
order parameter $q$ remains nonzero for any temperature, and thus
cannot be used to identify glassy freezing. Instead, we follow Mezard
and Young \cite{mezyoung}, and look for an instability to replica
symmetry breaking (RSB) within the paramagnetic phase. To do this, we
can set $m=0$, and note that the remaining equation for $\qab$ is in
fact {\em identical} to that describing the Sherrington-Kirkpatrick
model\cite{spinglass} in presence of random magnetic fields. This
model is also described with the Hamiltonian of Eq. (1), but this time
with $J_{ij}$'s being Gaussian random variables with zero mean and
variance $\langle J_{ij}^2\rangle\equiv {J^2\over zN}$, where $J$ is
the interaction of the original lattice model.

The advantage of mapping our equations to those of an infinite range
model will be also used in sections IV and V, where we examine the low
temperature structure of the glass phase.  In addition, the
calculation for obtaining the replica symmetry breaking (RSB) boundary
can be carried out using standard methods\cite{spinglass,AT},
following the approach of De Almeida and Thouless\cite{AT} (AT).  A
more general strategy for performing the RSB stability analysis, which
is applicable for arbitrary lattices, and to higher order fluctuation
corrections will be presented and discussed in section V. Here we
just quote the result valid to the leading nontrivial order, which
takes the form
\bea
1&=&{(\beta J)^2\over z}\int{Dx
\cosh^{-4}\left[\left((\b H_{RF})^2+{(\b J)^2q\over z}\right)^{1/2} x\right]
}\label{eq:SGBethe}.
\eea

As expected, in the large coordination ($z\rightarrow\infty$) limit,
the glass transition temperature vanishes, and our results reduce to
standard BWMFT predictions. For finite $z$, the SG phase emerges, in
agreement with the results of Mezard and Young \cite{mezyoung}. The
above equation can easily be solved numerically, and the results are
shown in Fig. 1(b). In contrast to the predictions of the SCS
approach, our SG phase emerges only for sufficiently strong random
fields. This is a simple result of the fact that in simple mean-field
treatments such as ours, different phases do not directly affect
each other, since there is no ``precursor'' of the ordering
in the disordered phase.

\subsection{Limit of large random fields}

It is interesting to note that the situation is especially simple in
the limit of large random fields. In the extreme case
$H_{RF}\rightarrow\infty$, the interactions can be ignored and all the
spins tend to align with their local random fields. In this case
there is only one thermodynamic state, so that the emergence of a
multitude of metastable states associated with the RSB instability is
clearly suppressed. We conclude that the glass transition temperature
must be depressed to zero as $H_{RF}\rightarrow\infty$, just as in
the case of spin glasses in uniform external fields first discussed
by de Almeida and Thouless. However, the asymptotic form of
$T_G (H_{RF})$ proves to be different in our case, which
may be significant for several experimental systems.

In the limit of large random fields, our results simplify
considerably, since the RSB boundary resides at very low temperatures.
The leading order behavior can be obtained by replacing the
$q(T)$ in Eq. (11) by its zero temperature limit $q(0)=1$,
and we find
\be
T_G\approx{4J^2\over 3zH_{RF}}\sim 1/H_{RF}.
\ee

It is interesting that our glass transition temperature thus decreases
very {\em slowly} with the random field strength, in contrast
to the form of the de Almeida-Thouless line \cite{AT,spinglass}(infinite ranged
spin glasses in a uniform field) where $T_G\sim\exp(-H^2/2J^2)$. This
fact could be particularly significant for strongly disordered
electronic systems \cite{tedrf,pastor}, where it would suggest the
possibility to observe the glassy behavior of electrons at finite
temperatures. The physical reason for this behavior in our case is not
obvious, but we will see that it reflects some very subtle features of
the low temperature glass phase, which will be discussed in section
IV.

From the physical point of view, an instability to RSB, such as we
find, is known\cite{spinglass,AT} to describe the emergence of an
extensive number of metastable states, and the associated slowing down
in the relaxational dynamics of the spins. Experimentally, this
results in the onset of the history dependence of cooling, and the
related bifurcation of field-cooled (FC) and zero-field cooled (ZFC)
spin susceptibilities.

\begin{figure}[h]
\bc \epsfxsize=4in \epsfbox{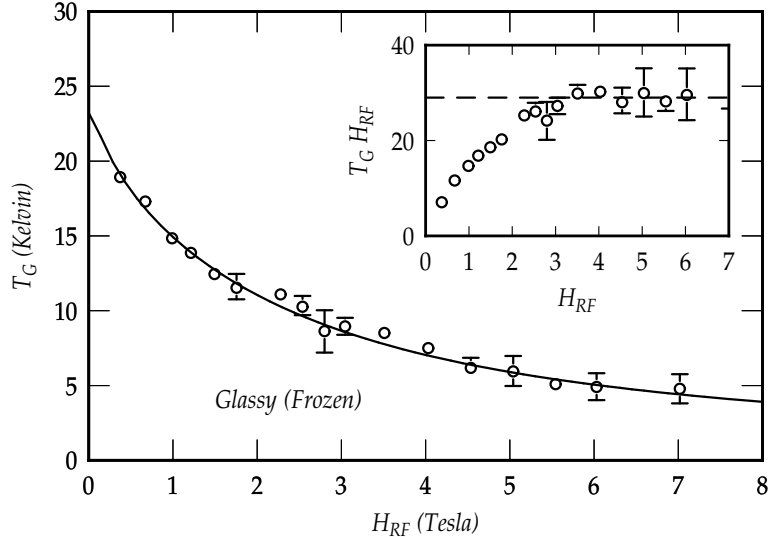}\ec \caption{Glass
transition temperature as a function of the random field strength.
Open circles are the experimental data for
$Fe_{0.31}Zn_{0.69}F_{0.2}$ (from Ref. [2]), and the line is the
prediction of Eq. (11). The inset shows how $T_G \sim 1/H_{RF}$ at
large fields. }\label{fig:montenegro}
\end{figure}
Such experiments have been carried out on diluted
antiferromagnets in uniform external fields, which have long been
believed to be realizations of the RFIM.  Here, the effective strength
of the random field can be varied by modifying the magnitude of the
external magnetic field.  In one such experiment,\cite{monten} the
field dependence of the ``irreversibility line'' has been determined,
defined as the temperature where the FC and ZFC susceptibilities start
to differ.  Interestingly, these experiments show a rather slow
decrease of this glass transition temperature, in agreement with our
predictions. We have digitized the data from Ref. [2], and compared
them to our predictions, as shown in Fig. 2, where an apparent
confirmation of our $T_G \sim 1/H_{RF}$ law can be seen (see
inset). While this agreement of our theory and experiment is
encouraging, it should not be taken too seriously, given the
uncertainties of the precise correspondence of the experimental system
and the RFIM that we consider.  More experiments on other related
systems would be welcome to test our predictions in more detail.

\section{General Lattices:  Legendre Transform Approach}

So far, we have seen how leading corrections to the BW theory can
be obtained on the Bethe lattice,  resulting in  the emergence of
the glass phase. This example was useful, because it automatically
introduces the correct order parameters, thus allowing the
emergence of the glass phase.  However, we would like to formulate
a more general approach, in order to demonstrate the generality of
our conclusions, and also to be able to systematically examine
higher order fluctuation corrections. In the case of the RFIM, the
desired order parameters cannot be introduced as for standard
spin-glass models, where one decouples the disorder-averaged
interaction term using a Hubbard-Stratonovich transformation. In
absence of random interactions, the ``bare'' disorder-averaged
Hamiltonian has only terms linear in the interaction $J$, but
higher order terms can be generated by fluctuations. In this case,
one has to introduce the order parameters ``by hand'', and then
expand the free energy to the lowest nontrivial order in the
interaction $J$, in order to obtain glassy ordering. To do this,
we follow a strategy introduced by Plefka \cite{plefka} and
Georges et al. \cite{georges}, and use a Legendre transform
approach, introducing external source fields $\xi_i^{\a\b}$ in
order to fix the Edwards-Anderson (EA) order parameters
$q_i^{\a\b}$. Here and in the rest of the paper, we are mostly
interested in the emergence of the glassy phase of the RFIM, so we
concentrate on the non-magnetic ($m=0$) solutions. The
disorder-averaged Helmholtz free energy takes the form
\be
-\b F = \lim_{n\rightarrow 0} \frac{\partial}{\partial n}
\left[-\b F_n\right],
\ee
where
\be
-\b F_n =
\ln \Tr\exp[{\b J\over z}\suma\sumij\sia\sja+
 \frac{(\beta H_{RF} )^2}{2}
\sum_i(\suma\sia)^2+\sum_i\sumalb\xiab_i \sia\sib],
\ee
The EA order parameters are given by
\be
\qab_i = {\partial(-\b F) \over \partial\xiab_i} = \langle \sia\sib \rangle ,
\ee
and the corresponding Gibbs free energy is
\be
-\b G = \lim_{n\rightarrow 0} \frac{\partial}{\partial n}
\left[-\b G_n\right] ;\;\;\;\;
-\b G_n  =-\b F_n  -
\sum_i\sumalb\xiab_i\qab_i.
\ee

\subsection{Extended mean-field theory}

To obtain an extended mean-field theory, we can expand $-\b G$ to a
desired order in powers of the reduced interaction $\ve\equiv \b J$,
while fixing the value of the independent variables $q_i^{\a\b}$.  In
doing this, one has to keep in mind that source fields $\xiab_i$ are
by Eq. (15) implicit functions of the order parameters $\qab_i$. The
{\em form} of these functions is also a function of $\b J$, so in
order to be consistent, one has to expand the fields $\xiab$ to a
given order in $\b J$ as well, while fixing $\qab_i$-s.  Defining $-\b
G_n /N \equiv g$, we can write
\bea
g (\ve )&=&g_o + g_{int} (\ve ),\\
g_{int} (\ve )&=& \sum_{k=1}^{\infty} \frac{\ve^k}{k!}\; g_k.
\eea
In this expression, the coefficients $g_k$ are functions of
$\qab$ evaluated at $\ve=0$.
Explicitly, we find
\be g_o [q] = f_o [\xi] -
\1n\sum_i \sum_{\a <  \b} \xiab_i \qab_i,\ee
where
\be f_o [\xi] = \1n\ln \Tr \exp \{-L_o \},\ee
and the reference effective action is defined by
\be L_o = -
 \sum_i\sum_{\a <  \b } \left( \xiab_i +
(\b H_{RF} )^2 \right)\sia\sib.\ee
The interaction terms take the form
\bea g_1 &=& \1n \langle \phi \rangle _o\\
     g_2 &=& \1n \langle \phi (\phi -\langle \phi \rangle _o +\chi - \langle \chi \rangle _o )\rangle _o,\\
&\vdots&\nn \eea
where we have defined
\bea
\phi &=& \frac{1}{z} \sum_{\a} \sum_{\langle ij\rangle }S_i^{\a}S_j^{\a}, \\
\chi &=& \sum_i \sum_{\a <\b} S_i^{\a}S_i^{\b}\frac{\partial}{\partial\ve}
\xiab_i .\eea
In these expressions, the averages are taken with respect to the
reference effective action $L_o$, which is a function of the
external fields $\xiab_i$. Note that since the coefficients $g_k$
are evaluated at $\ve=0$, after the cumulants are evaluated, the
external fields should be set equal to $\xiab_i (\ve =0 )\equiv
\xiab_o$, which are implicit functions of the independent
variables $\qab$, as determined by the following constraint
condition
\be \left. \qab =
\frac{\partial}{\partial \xiab_i} f_o [\xi ]\right|_{\xiab_i = \xiab_o}
= \; \langle S_i^{\a}S_i^{\b} \rangle _o .\ee
At this stage, we have taken the order parameter $\qab$ to
be uniform, since the system is translationally invariant after
averaging over disorder.

The equation of state is obtained from the saddle-point condition, giving
\be 0=\frac{\partial g[q ]}{\partial \qab} =
\sum_{\g\d}[\tilde{\partial}_{\g\d}
 f_o -q_{\g\d} ]\dab\xi^{\g\d}_o -\xiab_o +\dab\; g_{int},\ee
where we have used the notation $\dab \equiv \partial /
\partial \qab $, and $\tdab \equiv \partial /
\partial \xiab $. Using the constraint condition Eq. (26),
we can eliminate $\xiab_o$ and write
\be \qab = \tdab f_o [\partial \; g_{int} ].\ee
The described expansion can be carried out to any desired order in
$\ve$, generating fluctuation corrections to the Gibbs potential.
It is worth noting that, since this expansion is a series in
powers of the interaction $J_{ij}$, it generates corrections
due to {\em short-ranged} fluctuations, in contrast to the
usual loop expansion that is dominated by the long-wavelength
fluctuations. Accordingly, we expect that the prediction for the critical
behavior of the order parameter and response functions to maintain
the mean-field character if the expansion is truncated at any finite order.

\subsection{Leading fluctuation corrections: the $J^2$ theory}

The leading fluctuation corrections to the BW theory are obtain by
retaining terms up to second order in $\ve$, which is the lowest order
to which we can identify a glass phase. Using the up-down symmetry of
the Hamiltonian, we find (Appendix A) that $\langle \phi \rangle _o =
0$ and $\chi (\ve =0 )=0$, so that the above expressions simplify
giving 
\bea g_1 &=& 0\\
     g_2 &=& \1n \langle \phi^2 \rangle _o = \frac{1}{2z}[n + 2\sum_{\a < \b} \qab\qab] .
\eea

This expression is valid to any order in $\ve$, but using
in the leading theory (terms to $O(\ve^2 )$), we find
\be
\dab g_{int} = \frac{\ve^2}{z} \qab,\ee
giving
\be \qab = \tdab f_o [\frac{1}{z} (\b J)^2 q ].\ee
This equation of state is precisely the same as the $m=0$ limit of the
expression (Eq. (9)) that we have obtained on the Bethe lattice.  We
have thus demonstrated that to leading nontrivial order beyond the BW
theory, we find the same glassy phase for arbitrary lattices.  To this
order, which corresponds to introducing the Onsager reaction field
correction, the results are independent of the lattice geometry, as
also found in other problems \cite{spinglass,georges}.

\section{The $T=0$ glass phase: self-organized criticality}

The nature of the glass phase is in general much more complicated
than the usual ordered phases, and is characterized by a large
density of low energy excitations and low-lying metastable states.
Within the class of mean-field models for spin glasses, the structure
of the ordered phase has been investigated most extensively in the
vicinity of the glass transition, where the Parisi theory is
analytically tractable. Much less is known about the low temperature behavior,
although some of the most interesting phenomena are most pronounced there,
including the hysteresis and avalanche behavior.

\subsection{Hysteresis and avalanches behavior in presence of random fields}

To investigate the glass phase, we concentrate on the $T=0$
behavior of our model. To do this, we limit our attention to the
$J^2$ theory described above, where the RFIM maps to the SK
spin-glass model in presence of additional random magnetic fields.
This mapping allows us to use different methods to investigate the
$T=0$ behavior, which would be difficult to address using the
Parisi theory. In particular, we investigate the hysteresis and
avalanche behavior within our extended mean-field theory of the
RFIM, by following recent work of Pazmandi, Zarand and Zimanyi
(PZZ) \cite{zimanyi}. In their original calculation, PZZ have
examined the standard SK model, and the only modification that we
introduce is the additional presence of random magnetic fields.
The procedure that we follow is as follows. One first introduces a
large external magnetic field (in addition to fixed random
fields), so that all the $N$ spins are aligned with it. The
external field is then slowly reduced and the stability of the
spin configuration is examined with respect to any single spin
flips. As soon as the system becomes unstable, the spin
configuration is allowed to relax to a local energy minimum,
causing an ``avalanche''. The procedure is then repeated,
resulting in the system following a hysteresis loop. If the
external field is sweeped to large negative values (such that all
the spins align in the negative direction), and then the procedure
reversed, then the state of the system follows a ``major''
(external) hysteresis loop. If the field is instead reversed
before the major loop is completed, then the system embarks on one
of the minor hysteresis loops.  This procedure is illustrated is
Fig. 3 where a typical ``hysteresis spiral'' is presented.

\begin{figure}[h]
\bc \epsfxsize=4in \epsfbox{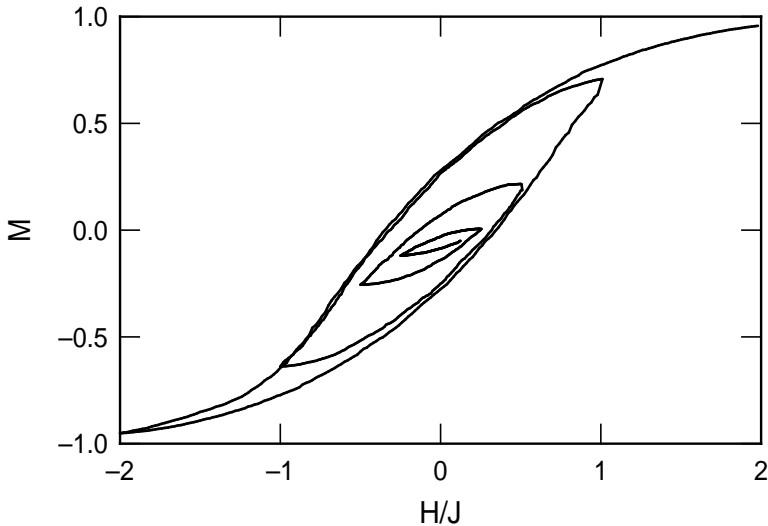}\ec \caption{A typical
hysteresis spiral.}
\end{figure}

\subsection{Distribution of local fields}

To characterize spin-glass state, we examine the probability
distribution $P(H_i)$ of local magnetic fields $H_i = h_i +
\sum_{j} J_{ij}S_j$ acting on a given spin $S_i$. In the high
temperature phase, $P(H_i)$ has a simple Gaussian distribution. To
see this, we recall that above the glass transition, the replica
symmetry remains valid, so that the presence of the interaction
term in the local effective action (see Eqs. (9-10); we use $m=0$
here) simply renormalizes the effective distribution of random
fields, such that $H_{RF}^{eff} = \sqrt{ (H_{RF})^2 + J^2 q/z}$.
The distribution $P(H_i)$ therefore remains a simple Gaussian of
width given by $H_{RF}^{eff}$. From Parisi theory\cite{spinglass},
we expect this distribution to acquire a nontrivial form upon
replica symmetry breaking, but we would like obtain its specific
form in the spin-glass state.

To calculate this quantity on the hysteresis loop, we follow a
simulation procedure identical to that used for the SK model by
Pazmandi, Zarand, and Zimanyi (PZZ)\cite{zimanyi}. In this
procedure, one considers a finite size system of $N$ spins, with a
given realization of random interactions $J_{ij}$ and random
fields $h_i$, and let the system explore the meta-stable states
sampled on the hysteresys loop. The values of the local fields
$H_i$ are then computed for every spin, and the procedure is
repeated for many realizations of disorder in order to generate a
large ensemble from which the desired distribution can be
computed. To implement this procedure, we have carried large-scale
simulations using systems with up to $N=3200$ spins, and obtaining
ensembles of $M=500,000$ data points from which the distribution
histograms were obtained. The resulting distributions for several
values of the random field strength are shown in Fig. 4. We find
that the distribution is characterized by the emergence of a
universal pseudogap of the form
\be P(H_i) \approx C\; H_i^{\a},\;\; (H_i << J) \ee
where $\a =1$ and $C=z/J^2$ is independent of the random field
strength. This universality is consistent with the findings of
PZZ\cite{zimanyi}, who have shown that it reflects the
self-organized criticality which characterizes the $T=0$ behavior
of the mean-field glassy systems. Our results confirm that the
conclusions of PZZ remain valid in presence of random fields, thus
representing a very robust property of glassy phases, at least
within the confines of the considered mean-field descriptions.

\begin{figure}[h]
\bc \epsfxsize=4in \epsfbox{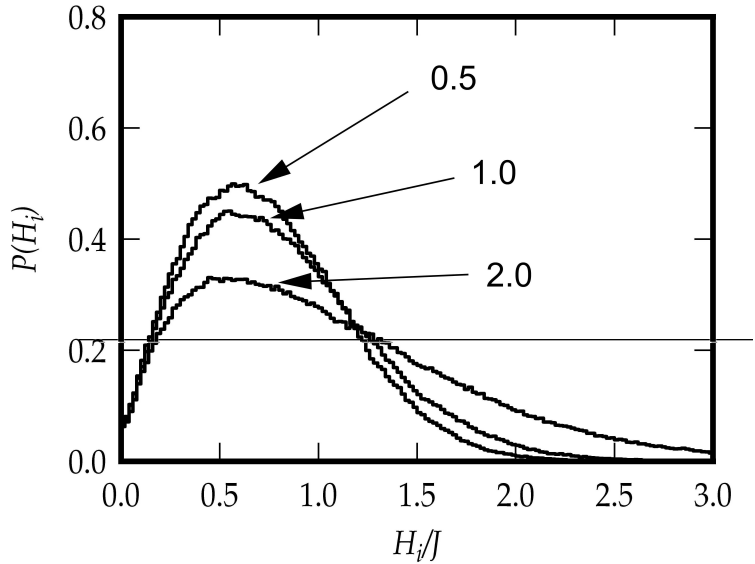}\ec \caption{Distribution
of local fields for N=3200, as a function of random field
strength, for $H_{RF}\sqrt{z}/J =$ 0.5, 1, 2. Note the universal
form of the pseudo-gap. The finite value of $P(0)\sim 1/\sqrt{N}$
is a finite size effect\cite{zimanyi}. }
\end{figure}

By following these procedures, we have carefully verified that all
the findings that PZZ have established\cite{zimanyi} for the SK
model also hold in presence of random fields, thus also apply to
the RFIM within the present mean-field formulation. In particular,
we have confirmed that the avalanche sizes are distributed on all
scales, and are characterized by a power-law distribution,
characteristic of {\em self-organized criticality}. It is most
remarkable that this critical nature is not confined to the ground
state, but persists for all the metastable states within the
hysteresis loop.  It is particularly interesting, as we have
explicitly verified by simulation, that not all local minima of
the energy have this property. Instead, the critical states form a
{\em subset} of metastable states that can be reached by the
described hysteresis procedure.  In this way, the ground state
seems not to have any special features, but rather to share the
same properties with an extensive number of critical metastable
states. This notion offers a natural origin for the criticality
that is found, since all the states along the hysteresis loop can
be considered to be {\em on the brink of an avalanche}, and are
therefore inherently unstable to weak perturbation, making the
criticality possible. In fact, it is precisely the requirement for
marginal stability of the hysteresis states that was used by PZZ
\cite{zimanyi} to derive the universal form of the local field
distribution. This argument examines the modification of the local
fields $H_i$ upon flipping a given set of $n_{flip}$ other spins.
This is given by
\be \delta H_i = H_i' - H_i +2\; \sum_{j flipped} J_{ij} S_j.\ee
One then computes the probability that the local field is
reversed, so that instabilities are created, triggering
avalanches.  Note, however, that the variations $\delta H_i$ is
{\em independent} of the value of the (external) random fields
$h_i$ present in our case. As a result, the rest of the argument
goes as in PZZ\cite{zimanyi}, giving the above marginality
condition.

From the historical perspective, evidence of marginal stability of
the spin glass phase in mean-field models has long been
appreciated based on Parisi and Thouless-Anderson-Palmer (TAP)
theory\cite{spinglass}. In addition, stability arguments requiring
$\a \ge 1$ have been presented by Palmer and
Pond\cite{palmerpond}, based on early ideas of
TAP\cite{spinglass}. However, it was not clear why this bound has
to be satisfied, or that even the prefactor $C$ assumes a
universal value. In this sense, the hysteresis and avalanche study
of PZZ has provided an important conceptual advance, linking this
universality with the self-organized criticality of the mean-field
models.

\subsection{Self-organized criticality and the AT line}

We have seen that the glass transition temperature in our
random-field case decreases very slowly, as $T_G \sim 1/H_{RF}$ at
large random fields.  This is very different than in the case of
the familiar AT line\cite{AT} of the SK model in a uniform field,
where the RSB temperature decreases {\em exponentially} with the
uniform field strength. In the following we present a simple
heuristic argument that explains the physical origin of both
behaviors based on the self-organized criticality of the glassy
ground states in these mean-field models.

What emerges from the analysis of PZZ for the SK model, and which
also applies in our extended mean-field formulation for the RFIM,
is the phenomenon that at $T=0$ the distribution of local fields
assumes a universal form at $H_i \rightarrow 0$, as given by Eq.
(33).  We have established this property for all the states within
the hysteresis loop following the methods of PZZ. However, it is
very likely that a similar universal distribution of local fields
also characterizes the exact ground state of the system. In fact,
this possibility has been proposed a long time ago both by TAP
\cite{spinglass} and by Palmer and Pond \cite{palmerpond},
consistent with the notion of the marginal stability of the spin
glass state for infinite-range models. Similarly to those of PZZ,
the arguments of Palmer and Pond can also be extended to include
the addition of random magnetic fields, resulting in a stability
bound that remains universal, i. e. independent of the random
field strength. To establish more firmly this universal form of
the pseudogap for the ground state, one would have to carry out
more elaborate numerical simulations. Efficient optimization
methods needed for such computations are available
\cite{optimization}, but this requires extensive efforts which are
beyond the scope of this paper.

For our purposes, we will assume that the pseudogap retains a
universal form for all $T=0$ critical states, including the
hysteresis states as well as the  ground state. If this is true,
we can now make a simple estimate of the ``condensation energy''
gained by glassy ordering, which should scale with the gap size.
Having in mind that the replica symmetric (high temperature)
distribution $P_{RS} (H_i)$ is a Gaussian of width given by
$H_{RF}^{eff}$, and the fact that the pseudo-gap has a universal,
linear form, we conclude that the gap size should scale as
\be E_{gap} \sim P_{RS} (0) \sim 1/H_{RF},\ee
at $H_{RF}\rightarrow \infty$. Here, we have used the fact that
$q(T=0)=1$, so to leading order $H_{RF}^{eff}\approx H_{RF}$ for
large random fields. Using this result, we can immediately
estimate the glass transition temperature, which should also scale
with this gap size, since $E_{gap}$ is the only energy scale
characterizing the ground state. We thus conclude that $T_g \sim
1/H_{RF}$, in agreement with our analytical calculations based on
the RSB analysis. It is worth emphasizing here that the simple
relationship between the gap energy and the random field strength
directly follows from the universality of the pseudo-gap form. Has
the gap had another functional form (e. g. a randomness-dependent
exponent $\alpha$), this relationship would be modified, resulting
in a different dependence of $T_g (H_{RF})$.  In this way, the
agreement between out analytical RSB results and the presented
heuristic arguments provides additional evidence for the
self-organized critical nature of the ground state in these
models.

Finally, it is interesting to examine how the presented heuristic
argument applies to the usual SK model in a uniform field. In this
case, the only modification is introduced by the fact that in the
high temperature phase, the uniform field simply shifts the entire
Gaussian distribution of local fields. As a result, $P_{RS}(H_i
=0,H) \sim \exp\{-H^2 /2J^2\}$, and we find the AT temperature
decreases exponentially with uniform field, in agreement with the
analytical results \cite{AT} of AT.

\subsection{Relevance for the Coulomb glass problem}

The nontrivial nature of mean-field glass models may be
particularly significant for the Coulomb glass problem,
relevant for disordered insulators. The Hamiltonian for
the Coulomb glass is given by
\be H = \sum_{ij}\frac{e^2}{r_{ij}}( n_i -\langle n\rangle )
(n_j - \langle  n_j \rangle ) -\sum_i \varepsilon_i n_i. \ee
Here, $n_i =0,1$ are the occupation numbers, $r_{ij}$ is the
inter-site distance, and $\varepsilon_i$ are the random site
energies. The transformation $S_i = 2 n_i -1$ immediately maps
this Hamiltonian to an \textit{antiferromagnetic} RFIM with
long-range interactions. Note that the presence of long-range
antiferromagnetic interactions leads to considerably stronger
frustration than that in the ferromagnetic RFIM considered in this
paper, which is found only for sufficiently strong randomness.
These differences may be crucial in low dimensional systems,
possibly leading to low temperature glassy phase for the Coulomb
system even if a similar behavior is suppressed for the
ferromagnetic RFIM case. However, on the mean-field level, even
the standard RFIM displays displays such glassy ordering, the
character of which may be closely related to that of the Coulomb
glass.

The best established property of the Coulomb glass model is
the existence of ``Coulomb gap'', as predicted by Efros
and Shklovskii (ES)\cite{efros}. According to ES,
the electronic system would be unstable, unless the
single-particle density of states (which corresponds to
our local field distribution) has a pseudo-gap
of the form
\be \rho (\varepsilon ) = C(d) \varepsilon^{d-1}, \ee
where $C(d)$ is a universal constant in $d$ dimensions.  More
precisely, ES have examined the stability of the system with respect
to one-electron excitations, showing that the form of Eq. (37)
represents an {\em upper bound} for the density of states. If this
bound would be saturated, then the pseudo-gap would assume a universal
form, but no convincing arguments have been presented why this should
happen.  However, large-scale numerical studies\cite{schreiber} have
obtained results similar to the ES predictions, failing to produce any
evidence of the hard gap. Still, the reason for saturating the ES
bound has remained a mystery.

Another aspect of the Coulomb glass that has not been properly
clarified is the presence or absence of a finite-temperature glass
transition\cite{thglass}, and the nature of the glass phase.
In this respect, the main difficulty was the absence of an appropriate
order parameter that would allow identifying the transition. Since the
random site energies in this model play a role of random fields, the
situation is identical as in the usual RFIM, and the usual EA
order-parameter cannot be used\cite{claireyu}. As we have seen, the
transition at best can have a character of an AT line, which should be
most easily identified in changes of the dynamics as the temperature
is lowered. Some numerical evidence that such a dynamical transition
may exist has been obtained by using the ``damage-spreading''
algorithms\cite{vojta}, giving hints of ergodicity breaking below a
certain temperature. In addition, several experiments\cite{expglass}
have reported history-dependent transport and other glassy features in
disordered insulators.

Despite the successes of the ES theory, several basic questions remain
unanswered. Most importantly, it is not clear whether the emergence of
the universal ES gap is related to the possible low temperature glassy
state of the model. In this respect, the scenario that we have
presented for the RFIM provides an interesting possibility. It is
conceivable that, as in our mean-field theory, a glassy phase exists
below a well-defined transition temperature, which corresponds to the
emergence of a large number of metastable states. If this phase were
characterized by self-organized criticality, then the associated
marginal stability would naturally explain the saturation of the ES
bound, and the resulting universality of the Coulomb gap.  An an
interesting way to address these questions would consists of examining
the $T=0$ hysteresis properties of this model, following the
approaches of PZZ, but this direction will be pursued elsewhere.

\section{Higher Order Corrections and the $1/z$ Expansion}

So far, we have examined the leading-order corrections to the
BW theory, producing the glass phase. On general grounds,
one does not expect that higher-order fluctuations corrections
of a finite order in $J$ would produce qualitative modifications
in this mean-field context. Nevertheless, we shall examine
the next-to-leading terms, in order to assert the convergence
properties of such an expansion.

\subsection{ The $J^4$ theory}

In section III, we have already calculated the terms up the to order
$J^2$ in the expansion of the Gibbs free energy. Terms up to order
$J^4$ can be computed using the same procedures, where we use the
average up-down symmetry of the problem outside the ferromagnetic
phase. In contrast to the $J^2$ theory, these additional terms depend
on the specific form of the lattice. To be specific, we concentrate on
the hypercubic lattice in $d$ dimensions (so that $z=2d$).  After
lengthy algebra, the resulting contributions up to order $J^4$ take the
form
\bea
g_3 &=& 0,\\
g_4 &=& \1n\left[\langle \phi^4\rangle_o^c +
3\langle \phi^2\chi '\rangle_o^c \right].\eea
Here, $\chi' = \partial\chi /\partial\ve$, and the brackets
$\langle\cdots\rangle_o^c$ indicates that only connected diagrams,
which give non-vanishing contributions, should be retained. This
refers to diagrams obtained by representing the quantity $\phi$
(see Eq. (24)) by a bond, since it stems from the interaction
terms connecting two nearest neighbors on a lattice. While each
power of $\phi$ involves a sum over all possible embeddings of
such a bond, only diagrams consisting of a close loops of such
bonds produce non-vanishing results, due to the average up-down
symmetry of the problem. Four different classes of such diagrams
are shown in Fig. 5, corresponding to the $J^4$ contributions that
we consider. \pagebreak
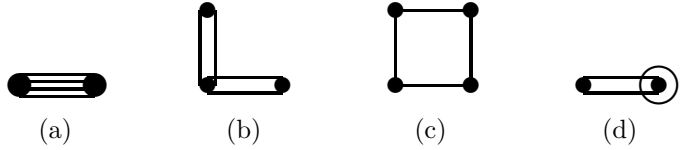
\begin{figure}
\begin{center}
\setlength\unitlength{1mm}
\begin{picture}(90,20)(-1,-1)
\thicklines \multiput(0,6.5)(0,-1){4}{\line(1,0){10}}
\multiput(0,5)(10,0){2}{\circle*{3}}
\multiput(25,5)(10,0){2}{\circle*{2}} \put(25,15){\circle*{2}}
\multiput(35,6)(0,-2){2}{\line(-1,0){10}}
\multiput(24,5)(2,0){2}{\line(0,1){10}}
\multiput(50,15)(10,0){2}{\circle*{2}}
\multiput(50,5)(10,0){2}{\circle*{2}}
\multiput(50,15)(0,-10){2}{\line(1,0){10}}
\multiput(50,15)(10,0){2}{\line(0,-1){10}}
\multiput(75,5)(10,0){2}{\circle*{2}} \put(85,5){\circle{5}}
\multiput(85,6)(0,-2){2}{\line(-1,0){10}} \put(2.5,-2){(a)}
\put(27.5,-2){(b)} \put(52.5,-2){(c)} \put(77.5,-2){(d)}
\end{picture}
\end{center}
\caption{Diagrams contributing to order $J^4$. The diagrams (a),
(b) and (c) correspond to the $\langle\phi^4\rangle_o^c$ term, and
(d) to the $\langle\phi^2\chi '\rangle_o^c$ term.  Here, the
$\phi$ bonds are represented by a full line, and the local $\chi
'$ term is represented by an open circle.}
\end{figure}
The evaluation of these terms is straigthforward.
Here, we
only emphasize the following properties of these diagrams, that are
valid even to higher order in $J$, as follows.
(1) Diagrams (a), (b) and (d) contain vertices with four emerging
bonds, that in the replica calculation give rise to expressions which
include moments of the form $r_{\a\b\g\d} = \langle
S_i^{\a}S_i^{\b}S_i^{\g}S_i^{\d}\rangle_o$, with $\a
\neq\b\neq\g\neq\d$. Such moments cannot be simply expressed in
terms of the order parameter $\qab$.  In contrast, the ``loop'' diagram (c)
is expressed as a power series involving only powers of $\qab$.
(2) The lattice embedding factors, which specify the $z$-dependence,
do not depend on the specific replica decorations that have to be
carried out on each diagram, but are determined only by the topology
of the diagram.
(3) We find that diagrams (b) and (d) result in identical expressions,
except for the lattice embedding prefactors, which are different. As a
result, cancellations occur, such that the sum of (b) and (d) produces
a contribution to the free energy which is of order $J^4 /z^3$, as is
the contribution of (a).
(4) The leading contribution, of order $J^4 /z^2$, follows only from the
loop diagram (c).
(5) The diagrams (a), (b), and (d) consist of self-retracting paths,
and as such are identical for both a hypercubic and the Bethe lattice
with the same $z$. The ``open loop'' diagram (c) is specific to the
hypercubic, but absent for the Bethe lattice.

In examining the $J^4$ (and higher order) corrections, we are
primarily interested in identifying a limit where these terms are
``small'', so that only the leading $J^2$ contributions may be
retained.  On a Bethe lattice, all the contributions are of order
$1/z^3$, thus can be ignored in the limit of large coordination. It is
easy to see that the same conclusion applies to arbitrary lattices
with purely random interactions such that $\langle J_{ij}\rangle =0$,
as found in spin glass models\cite{georges}, since in this case only
self-retracting paths survive. In contrast, for the RFIM on general
lattices, where the interactions are uniform, the ``open loop''
diagrams (c) survive, and provide the leading contribution, which is
of the same order as in the $J^2$ term. We expect that similar
conclusions apply to higher order contributions as well.  This
property, that the ``open loop'' diagrams provide a leading
contribution in large dimensions is well known\cite{metzner}, and has
been extensively used to study models of strong electronic correlation
in large dimensions\cite{dmfrmp}. To get the leading contributions for
large $z$, one would therefore have to sum up such open loop
contributions to all orders in $J$.  In absence of randomness such
re-summations have been carried out\cite{metzner,dmfrmp}, but at least
for the electronic models, the results were not qualitatively different
from those obtained on the Bethe lattice, where only the second order
terms may be retained.

In our case such re-summations are more difficult to carry out due to
the presence of randomness (replica indices), and will not be
attempted here. Instead, we will show that if one examines the {\em
joint} limit of large coordination and strong random fields, even the
$J^4$ terms represent sub-leading contributions, and the simple $J^2$
theory suffices. To show this, we only consider the leading loop
contributions (diagram (c)), which takes the form
\be
g_4 = \frac{3}{z^2}\left[
6 \sum_{\a\neq\b} (\qab )^2 +
4 \sum_{\a\neq\b\neq\g} \qab\qbg\qga +
\sum_{\a\neq\b} (\qab )^4 +
2 \sum_{\a\neq\b\neq\g} (\qab )^2 (\qbg )^2 +
 \sum_{\a\neq\b\neq\g\neq\d} \qab\qbg\qgd\qda \right] .\ee

\subsection{1/z corrections to the equation of state: RS solution}

A general expression for the equation of state, valid to arbitrary
order is given by Eq. (28).  To compute the relevant $J^4$ corrections
in the replica symmetric case, we calculate the variation of
$g_{int}$, which to this order reads
\be \left. \dab\;  g_{int}\;\right|_{RS} = \frac{\ve^2}{z} q +
3\left( \frac{\ve^2}{z}\right)^2 q(1-q)^2 ,\ee
where we have taken the $n\rightarrow 0$ limit.

We are especially interested in examining the form of the
solution in the large $H_{RF}$ limit, where the RSB transition occurs
at low temperatures. We examine the relative magnitude of the
terms appearing in $g_{int}$. At first sight, the $J^4$ terms seems to
be dominant in the low $T$ limit, since it is of order
$\ve^4  = (\b J )^4$. However, note that this term is also proportional
to $\d q^2 = (1-q)^2$, which is small at low temperatures,
since $q\rightarrow 1$ at $T\rightarrow 0$. To see this, it suffices
(to leading order) to compute $q(T)$ at $J=0$, giving
\be q^{J=0} = 1-\frac{2}{\sqrt{\pi}}\frac{1}{\b H_{RF}} \cdots .\ee
This gives $\d q\sim (\b H_{RF})^{-1}$, so that
$(\b J)^4 q(1-q)^2 \sim (\b J)^2 (H_{RF}/J)^{-2}$.
 The $J^4$ term is down by a factor $(H_{RF} /J)^{-2}$,
thus contributing only to {\em sub-leading} order in the limit
of large random fields.

\subsection{RSB stability analysis: a general approach}

Within the $J^2$ theory (sections II and III), we have mapped the RFIM to an equivalent
infinite-range model, making it possible to carry out the RSB
analysis similarly as for the SK model. In the following,
we present a general approach to the RSB stability analysis, which
can be used even if higher order terms are retained.

To identify the RSB instability we follow AT\cite{AT}, and examine the
variations of the Gibbs free energy with respect to the deviations
from replica symmetry $\qab = q + \delta\qab$. From the definition of
the Gibbs free energy, Eq. (19), the corresponding stability matrix is
\be \partial_{\a\b ,\g\d} \; g
=-\dab\xi^{\g\d} + \partial_{\a\b ,\g\d}\; g_{int}.\ee
Here, we have used the notation
$\partial_{\a\b ,\g\d} \equiv \partial^2 / \partial \qab \partial\qgd$.
To eliminate the quantity $\dab\xi^{\g\d}$, we take a variation of
the constraint condition  Eq. (26), and obtain
\be 0=\delta_{\a\b ,\g\d}
-\sum_{\mu\nu} \tilde{\partial}_{\g\d ,\mu\nu}\; f_o [\xi] \dab
\xi^{\mu\nu},\ee
where we have used
$\tilde{\partial}_{\a\b ,\g\d} \equiv \partial^2 /
\partial \xi^{\a\b} \partial\xi^{\g\d}$. To write these expressions in a more compact
form, we introduce a matrix notation
$(\hat{g}'')_{\a\b ,\g\d} \equiv \partial_{\a\b ,\g\d} \; g$;
$(\hat{g}''_{int})_{\a\b ,\g\d} \equiv \partial_{\a\b ,\g\d} \; g_{int}$;
$(\hat{\xi}')_{\a\b ,\g\d} \equiv \partial_{\a\b}\; \xi^{\g\d}$;
$(\hat{f}''_{int})_{\a\b ,\g\d} \equiv \tilde{\partial}_{\a\b ,\g\d}\;
f_o [\xi]$, giving
\bea \hat{g}'' = -\hat{\xi}' + \hat{g}''_{int}\\
0 = \hat{I}  - \hat{f}''\cdot \hat{\xi}'.\eea
In this form, the matrix $ \hat{\xi}'$ can be eliminated and we get
\be \hat{g}'' = - (\hat{f}'')^{-1} + \hat{g}''_{int}.\ee

To simplify the calculation further, we note that in the high temperature
phase all the eigenvalues of $ \hat{g}''$ are positive, but the RSB instability
is identified\cite{AT} when at least one of its eigenvalues vanishes,
such that the {\em determinant} (at fixed $n$) vanishes. Using this property,
an equivalent approach to the RSB stability analysis can be formulated by examining
the stability of an auxiliary matrix
\be \hat{g}''_1 = \hat{f}''\cdot \hat{g}''.\ee
This is true, since
\be \det(\hat{g}''_1)= \det(\hat{f}'')\det(\hat{g}''),\ee
and it can be verified by explicit calculation that the matrix $\hat{f}''$
remains nonsingular in the temperature range of interest. In other words,
to identify the RSB instability, we need to compute the eigenvalues of
the auxiliary stability matrix
\be \hat{g}''_1 = \hat{f}''\cdot \hat{g}''_{int} -\hat{I}.\ee

We now present a general strategy for computing the eigenvalues of this
matrix. Our first observation is that any replica matrix of the form
considered, has at most three different matrix elements, as discussed
by AT\cite{AT}. In addition\cite{AT}, for any such matrix there
can be found three different (degenerate) eigenvectors. Most remarkably,
the {\em form} of these eigenvectors {\em does not} depend on the
vale of the corresponding matrix elements. Thus, any replica
matrix has the same eigenvectors, and only the corresponding eigenvalues
depend on the value of the matrix elements. Since a product of two replica matrices
is again a replica matrix, we conclude that the relevant eigenvectors in our
case are identical to those computed by AT, and we only need to evaluate the
relevant eigenvalues. Only one of the eigenvalues vanishes at the RSB transition,
call it $\lambda_3$, and the corresponding eigenvector $\vec{x}_3$. The eigenvalue
$\lambda_3$ in our case can be computed by acting on $\vec{x}_3$ with the
matrix $ \hat{g}''_1$, and we find
\be \lambda_3 = \lambda_3^f\cdot\lambda_3^{int} -1,\ee
where $ \lambda_3^f$ and $\lambda_3^{int}$ are the respective
eigenvalues of the matrices $ \hat{f}''$ and $\hat{g}''_{int}$. We
emphasize that this strategy is not specific to our $J^4$ theory, but
is valid to arbitrary order in the expansion.

To obtain the desired $J^4$ corrections, we need to compute the
corresponding corrections to the matrices $ \hat{f}''$ and
$\hat{g}''_{int}$, as done in Appendix B.  For general $H_{RF}$,
we find that the $J^4$ corrections are of the {\em same} order as
the $J^2$ ones, leading to a glass transition temperature $T_G
\sim 1/\sqrt{z}$. Clearly, all the higher order terms coming from
the ``loop'' diagrams are also of the same order, and would have
to be included as well, in order to collect all the leading
contributions in a $1/z$ expansion, similarly as in other problems
in large dimensions\cite{dmfrmp}.

However, as shown in Appendix B the expressions simplify in the
limit of large random fields, where all the $J^4$ contributions
are down by a factor $(J/H_{RF})^2$, and to leading order
expressions are obtained by simply retaining only the $J^2$ terms.
We conclude that the $J^2$ theory, which provides the leading
nontrivial order in our extended mean-field theory, represents an
asymptotically exact formulation in the {\em joint} limit of large
coordination and large random fields. These conclusions have been
obtained by examining the example of a simple hypercubic lattice
with nearest neighbor interactions, in the limit of large
coordination. Since the general structure of the diagrammatics and
the relevant $z$-dependence is qualitatively the same for general
lattices, we expect these results to hold for any model with
short-range interactions.

The situation is more complicated for models with long-range
interactions, such as the Coulomb glass. In such cases, the
diagrammatic expansion cannot be truncated to any finite order in the
interaction, in order to avoid well-known divergences associated with
the screening processes\cite{reichl}. The simplest consistent
treatment has to sum-up all the ``chain'' diagrams, leading to the
Debye-Huckel approximation\cite{reichl}, which is also known as the
random-phase approximation in the electronic context\cite{fetter}.
This class of diagrams is, in fact, equivalent to the class of
``loop'' diagrams in our expansion of the free energy, which provides
the leading contributions in the limit of large coordination. Thus, to
address the behavior of the Coulomb glass model, one should extend our
calculation to sum up all the loop diagrams, which can
straightforwardly be done even within our formulation. The resulting
theory should be capable of addressing the interplay of screening
effects and glassy freezing, a topic of great relevance for disordered
electronic systems.

\subsection{Weak random fields and self-generated glass}

Another instance where terms not contained in the simplest $J^2$
theory may be important is the limit of weak randomness.  On
general grounds, one there expects the system at low temperature
to assume some uniform order. In many cases the corresponding
transition has a first-order character, so the upon rapid cooling
the system may remain trapped in a meta-stable state and undergo
glassy freezing. This process is believed to be even more likely
in presence of competing uniform interactions, which typically can
depress the uniform ordering down to very low temperatures. A
model for this behavior has been proposed a long time
ago\cite{teddave}, based on earlier work \cite{tdw} that
emphasized the relation between mode-coupling
theories\cite{goetze} of supercooled liquids and a special class
of infinite range spin-glass models displaying a first order glass
transition scenario.  The possibility of glassy freezing in
absence of randomness has recently attracted renewed attention,
and several studies\cite{self-glass,ioffe} have concentrated on
uniformly frustrated infinite-range models where these processes
can be studied in detail. However, a more general approach would
be even more useful, where one could examine the competition
between uniform and glassy orderings in a controlled scheme, and
which could be applied to models with realistic interactions and
lattice geometries.

In principle, these questions can be addressed within our approach
by examining the fate of the glassy phase in the limit of weak
random fields, an idea that was introduced a long time
ago\cite{teddave}.  In the simplest $J^2$ theory, and outside the
ferromagnetic phase our model maps to the SK model in presence of
random fields, leading to the glass transition line given in Fig.
1(b). As the random fields are reduced, the glassy phase is
enhanced, but for sufficiently weak randomness, glassy freezing is
preempted by uniform ferromagnetic ordering. On the other hand, if
we restrict our attention to the non-magnetic ($m=0$) solution,
then the glass transition line can be extended to $H_{RF}=0$,
leading to $T_G (H_{RF} =0)=J/\sqrt{z}$.  In this way, our
formulation may be considered the simplest approach that can lead
to glassy behavior in absence of randomness. However, the
prediction of this lowest order approximation cannot be considered
as reliable for weak random fields, since corrections to any order
in $J$ make contributions of the comparable magnitude, even in the
limit of large coordination.  In fact, using expressions that we
have obtained within the $J^2$ theory, it is not difficult to
compute the resulting correction to the glass transition
temperature, which remains of order $1/\sqrt{z}$, but with an {\em
increased} prefactor. The details will not be elaborated, since
stopping at any finite order in $J$ is clearly not sufficient. The
enhancement of the glass phase in the limit of weak random fields
due to these fluctuation corrections may indicate the possibility
that, once all the leading corrections are retained, the glass
transition would {\em precede} any uniform ordering as suggested
by the SCS theory of Mezard and Young\cite{mezyoung}. If this is
correct, it would indicate that the convergence of the $1/z$
expansion is not uniform as a function of the random field
strength, since $T_G (H_{RF} \rightarrow \infty)\sim 1/\sqrt{z}$,
but $T_G (H_{RF}\rightarrow 0) \sim O(1)$. In that case, the
correct mean-field theory should not be formulated by performing a
$z$-dependent rescaling of the interactions and then letting
$z\rightarrow\infty$. Instead, the formulation should retain all
the leading $1/z$ corrections, in finite dimensions. It is
interesting to note that recent work of Lopatin and
Ioffe\cite{ioffe} is closely related to the approach that we
propose, since it singles out all the leading $1/d$ corrections
for a specific uniformly frustrated model, in the limit of large
dimensions.  In this particular model, the interactions are
sufficiently frustrated, precluding any uniform ordering, and
resulting in a particularly simple large coordination limit. In
more general cases, competition between uniform and glassy
ordering and the possibilities of having either first or second
order transitions should be considered and may be determined by
the details of the interactions or the presence of disorder.

\section{Conclusions}

In this paper, we have presented a systematic approach that can
incorporate short-range fluctuation corrections to the standard
Bragg-Williams theory of the random-field Ising model. We have
shown that if the correct order parameters are introduced,
corrections to even the lowest nontrivial order immediately result
in the appearance of a glassy phase for sufficiently strong
randomness.  This low-order treatment is shown to be sufficient
for large randomness, where it provides the leading corrections in
the limit of large coordination. The structure of the resulting
glassy phase is very similar to that found in familiar infinite
range spin glass models, and is characterized by universal
behavior emerging from the self-organized criticality of the
ground state.

The major puzzle that remains to be resolved is the extent to
which the application of these mean-field ideas is relevant to the
low temperature behavior of low dimensional systems with short
range interactions. An alternative approach, based on droplet
arguments\cite{fisher} presents a very different scenario,
particularly in situations where external fields, either uniform
or random, explicitly breaks the symmetry of the Hamiltonian. In
this instance, droplet arguments would preclude the existence of
any finite-temperature glass transition, in contrast to the
mean-field predictions.  In addition, recent numerical results
\cite{middleton} on $d=3$  RFIM have  also been been used to argue
against the existence of a finite-temperature transition in a
field. In this context, it is worth noting that self-organized
criticality is {\em not} found in recent studies \cite{sethna} of
hysteresis and avalanche behavior of RFIM with short-range
interactions \cite{sethna} in low dimensions. For these models,
although hysteresis behavior is present, the distribution of
avalanche sizes is bounded, and criticality is found only by
fine-tuning the parameters of the system to a particular point of
the phase diagram. Similar results have been obtained in studies
that have examined the sensitivity of the $d=1,2$ RFIM to small
random perturbations of the quenched disorder \cite{rieger}.

In our opinion, more general emergence of self-organized
criticality similar to that found in  mean-field glassy models
most likely requires the existence of longer-range spin-spin
interactions and/or high spatial dimensions. On the other hand,
experiments\cite{magnets} measuring the Barkhausen noise on
several ``hard magnets'' have indicated that power-law
distribution of avalanche sizes and avalanche times, consistent
with self-organized criticality. Such behavior may be a result of
the fact that in such systems the dominant interactions have a
dipolar, thus longer-range character, bringing the behavior of
these materials closer to the predictions of mean-field glassy
models. These features may also be of particular importance in
applications of the RFIM to disordered electronic systems and the
related physics of the Coulomb glass behavior.

In this work, have also outlined how our theory could be extended
to examine models with either longer-range interactions or the
limit of weak random fields, which is of particular importance to
the long-puzzling question of the self-generated glassiness in
uniform systems. Our theory is closely related to other recent
approaches \cite{spinglass,mezyoung,self-glass,ioffe} that address
the emergence of glassy phases on a mean-field level. These
theories taken as a whole appear to shed light on a number of
experimentally relevant systems, and present a fairly complete and
consistent picture of glassy behavior.

\acknowledgements
We thank E. Dagotto, A. Moreo, T. R. Kirkpatrick, G. Kotliar,
A. E. Ruckenstein, G. Zimanyi, and G. Zarand for useful discussions.
This work was partially supported by the NSF grant DMR-9974311, the
National High Magnetic Field Laboratory, and the Alfred P. Sloan
Foundation (V. D.).

\appendix

\section{Cancellations due to up-down symmetry}

In investigating the glassy phase of the RFIM outside the FM
phase, and in zero uniform field, we can make use of the fact that
after averaging, the system respects up-down symmetry. As a
result, a number of terms vanish, so that the expressions
simplify. In the following, we discuss these cancellations in some
detail.

\subsection{Moments of spins on different sites}

The term $\langle \phi \rangle_o$  contains expectation values of
a product of spins on different sites. Since the averages are
computed at $\ve=0$, such terms very generally factor out, so that
we can write
\be \langle S_{i_1}^{\a_1}\cdots S_{i_k}^{\a_k}\rangle _o =
 \langle S_{i_1}^{\a_1}\rangle _o \cdots \langle S_{i_k}^{\a_k}\rangle _o = m^k,\ee
where $m$ is the magnetization. Outside the FM phase, and at
$H=0$, the magnetization and thus all such moments vanish.

\subsection{Evaluation of $\langle \phi \rangle _o$ and $\langle \chi \rangle _o$}

An immediate consequence of the above factorization property is
the fact that $\langle \phi \rangle _o =0$, since this expression
contains products of two spins on different sites.  The expression
for $\chi$ contains a derivative of the source field of the form
$\frac{\partial}{\partial\ve}\xiab_i$.  To compute this
derivative, we use the fact that due to the definition of the
Legendre transform, we can write
\be \xiab_i = \frac{\partial}{\partial\qab_i}\;( \b G )\ee
To calculate $\langle \chi \rangle _o$, we need to compute the
derivative of $\xiab_i (\ve )$, evaluated at $\ve=0$, and we get
\be \left. \frac{\partial\xiab_i}{\partial\ve}\right|_{\ve=0} =
\left. \frac{\partial}{\partial\qab_i}
\frac{\partial}{\partial\ve}\; (\b G )\right|_{\ve=0} =
-\frac{\langle \phi \rangle _o}{\partial\qab_i} =0. \ee We
therefore conclude that $\langle \chi \rangle _o =0$ as well.

\subsection{Terms with odd powers of $\phi$}

In evaluating higher order terms in the $\ve$-expansion, terms of
the form
$\langle \phi^{p}\psi (\chi ',\chi '')\rangle _o$
appear, where $\psi (\chi ',\chi '')$ is an arbitrary polynomial
function of $\chi '$ and $\chi ''$, and $p$ is an odd number. To
evaluate such terms,  we note that as before, spin moments on
different sites factor out, but we still have to compute
nontrivial spin moments of the form
$\langle S_i^{\a_1}\cdots S_i^{\a_r} \rangle _o $, with $r$ having
the same parity as $p$, i .e $r$ is odd.  Here, we have used the
fact that quantities $\chi ' = \frac{\partial}{\partial \ve}\chi$
and $\chi '' = \frac{\partial^2}{\partial \ve^2}\chi$ are {\em
quadratic in local variables}, i. e. contain products of the form
$S_i^{\a} S_i^{\b}$. In addition, the considered moments will have
an odd number of spins only if the considered lattices have no
odd-membered rings, such as found for example on a triangular
lattice. If any of the replica indices coincide, then $(S_i^{\a}
)^2 =1$, and an {\em even} number of spins drop out, but the
remaining expression still has the form
\be M_s = \langle S_i^{\a_1}\cdots S_i^{\a_s} \rangle _o, \;\;\;
\a_1 <\a_2 <\cdots \a_s .\ee
Using well-know properties of replicas \cite{spinglass}, it is
readily seen that
\be M_s = \overline{\langle S_i \rangle ^s} = \overline{m^s}=0.
\ee Therefore the expression of the considered form vanish as well
for lattices with no odd-membered rings.

\section{variations of the Gibbs free energy}

A general procedure needed to obtain the equation of state and the RSB
stability criterion involves computing the variations of the Gibbs
free energy with respect to the order parameter $q^{\a\b}$. In the
following we outline how these variations can be computed by concentrating
only on the leading-order contributions from the ``loop'' diagram.

\subsection{Calculation of $\dab g_{int}$}

The calculation of $\dab g_{int}$ to order $J^2$ is already computed in
section III A. here, we compute the $J^4$ correction.  Defining the
matrix $\hat{A}=\hat{I}+\hat{Q}$, we can write
$g_4 = \frac{3}{z^2} \Tr[\hat{A}^4]$, and find
\be \dab \; g_4  = \sum_{\g\d} \frac{\partial g_4}{\partial A_{\g\d}}
\frac{\partial A_{\g\d}}{\partial \qab}\
= \frac{24}{z^2}(\hat{A}^3)_{\a\b}.\ee
In the RS limit $\qab =q$, and taking $n\rightarrow 0$ this reduces
to the expression of Eq. (41).

\subsection{Calculation of   $\hat{g}_{int}''$}

From Eq. (31), the $J^2$ contribution reads
\be \partial^2_{\a\b ,\g\d}\; g_{2} = \frac{2}{z} \d _{\a\b ,\g\d}.\ee
The $J^4$ contribution can be calculated using the same procedure as
for the first variation, and we find
\be \partial^2_{\a\b ,\g\d}\; g_{4} =
\frac{48}{z^2}\left[ \delta_{\a\g}\sum_{\mu}A_{\b\mu}A_{\mu\d}
+A_{\a\g}A_{\b\d}
+\delta_{\b\d}\sum_{\mu}A_{\a\mu}A_{\mu\g}\right] .\ee
In the replica symmetric theory, the resulting matrix elements
of $\hat{g}_{int}''$ are given by
\bea \tilde{P}&=& (\hat{g}_{int}'')_{\a\b ,\a\b} =
\frac{\ve^2}{z} + \frac{1}{2}\left(\frac{\ve^2}{z}\right)^2
(2(1-q^2 )+1),\nne
     \tilde{Q}&=& (\hat{g}_{int}'')_{\a\b ,\a\d} =
\frac{1}{2}\left(\frac{\ve^2}{z}\right)^2
(2q(1-q )+q),\nne
     \tilde{R}&=& (\hat{g}_{int}'')_{\a\b ,\g\d} =
\frac{1}{2}\left(\frac{\ve^2}{z}\right)^2 q^2.\eea
Given these matrix elements, one can immediately evaluate
the relevant eigenvalue\cite{AT} of the matrix $\hat{g}_{int}''$,
which takes the form
\be \lambda_3^{int} = \tilde{P}-2\tilde{Q}+\tilde{R}
=\frac{\ve^2}{z} +
\frac{3}{2}\left(\frac{\ve^2}{z}\right)^2 (1-q)^2.\ee
In this expression, we note that similarly as in the
computation of the $J^2$ correction to the RS equation of state,
the $J^4$ correction is proportional to $\d q=1-q$, and is
therefore down by a factor $(J/H_{RF})^2$  compared to the
leading term. We again conclude that in the limit of large
random fields, to leading order it suffices to retain the $J^2$
contribution.

\subsection{Calculation of $\hat{f}''$}

The functional $f_o [\xi]$ is the free energy of free spins in presence
of fields $\xiab$, and we are interested in computing its second
variation at the saddle point where $\xiab =\dab g_{int}$. Therefore,
this evaluation is almost identical as for the SK
model\cite{spinglass}. In addition, since this quantity is evaluated
in the RS theory, $\dab g_{int}$ assumes the form that we have already
discussed when we examined the RS equation of state. We conclude
that for this quantity as well, to leading order in the limit of large
random fields, the argument of $\hat{f}''$ can be replaced by its $J^2$
approximation, and we straightforwardly obtain Eq. (11).

\end{document}